\documentclass[conference]{IEEEtran}
\usepackage[utf8]{inputenc}
\usepackage{siunitx}
\usepackage{graphicx}
\usepackage{tikz}
\usepackage{comment}
\usetikzlibrary{shapes, arrows, calc, positioning, fit}
\usepackage{pgfplots}
\pgfplotsset{compat=1.14}
\usetikzlibrary{patterns}
\usepackage{url}

\title{FastFabric: Scaling Hyperledger Fabric to 20,000 Transactions per Second}
\author{\IEEEauthorblockN{Christian Gorenflo}
\IEEEauthorblockA{
University of Waterloo\\
Waterloo, Canada\\
Email: cgorenflo@uwaterloo.ca}
\and
\IEEEauthorblockN{Stephen Lee}
\IEEEauthorblockA{University of Massachusetts\\
Amherst, USA\\
Email: stephenlee@cs.umass.edu}
\and
\IEEEauthorblockN{Lukasz Golab, S. Keshav}
\IEEEauthorblockA{
University of Waterloo\\
Waterloo, Canada\\
Email: [lgolab, keshav]@uwaterloo.ca}}

\date{December 2018}

\begin{document}

\maketitle

\begin{abstract}

Blockchain technologies are expected to make a significant impact on a variety of industries. However, one issue holding them back is their limited transaction throughput, especially compared to established solutions such as distributed database systems.  In this paper, we re-architect a modern permissioned blockchain system, Hyperledger Fabric, to increase transaction throughput from 3,000 to 20,000 transactions per second. We focus on performance bottlenecks beyond the consensus mechanism, and we propose architectural changes that reduce computation and I/O overhead during transaction ordering and validation to greatly improve throughput.  Notably, our optimizations are fully plug-and-play and do not require any interface changes to Hyperledger Fabric.

\end{abstract}

\section{Introduction}

Distributed ledger technologies such as blockchains offer a way to conduct transactions in a secure and verifiable manner without the need for a trusted third party.  As such, it is widely believed that blockchains will significantly impact industries ranging from finance and real estate to public administration, energy and transportation~\cite{espinel2015deep}.  However, in order to be viable in practice, blockchains must support transaction rates comparable to those supported by existing database management systems, which can provide some of the same transactional guarantees.

In contrast to permissionless blockchains, which do not restrict network membership, we focus on \emph{permissioned} blockchains, in which the identities of all participating nodes are known.  
Permissioned blockchains are suitable for many application domains including finance; e.g., the Ripple\footnote{\url{https://ripple.com}} blockchain aims to provide a payment network for currency exchange and cross-bank transactions akin to the current SWIFT system.  

From a technical standpoint, we observe an important distinction between these two types of blockchains.  The trustless nature of permissionless blockchains requires either proofs of work or stake, or expensive global-scale Byzantine-fault-tolerant consensus mechanisms~\cite{vukolic2015quest}. Much recent work has focused on making these more efficient.  On the other hand, permissioned blockchains typically delegate consensus and transaction validation to a selected group of nodes, reducing the burden on consensus algorithms.  While even in this setting consensus remains a bottleneck, it is being addressed in recent work~\cite{sousa2018byzantine, yin2018hotstuff}, motivating us to look beyond consensus to identify further performance improvements.

In this paper,
we critically examine the design of Hyperledger Fabric~1.2 since it is reported 
to be the fastest available open-source permissioned blockchain~\cite{Dinh2017d}.  
While there has been some work on optimizing Hyperledger Fabric, e.g., using aggressive caching \cite{Thakkar2018}, we are not aware of any prior work on re-architecting the system as a whole\footnote{Please
refer to Section \ref{sec:related} for a detailed survey of related work.}.
We hence design and implement several architectural 
optimizations based on common system design techniques that together improve the end-to-end transaction throughput 
by a factor of almost 7, from 3,000 to 20,000 transactions per second,  while decreasing block latency.  Our specific contributions are as follows:
\begin{enumerate}
    \item \emph{Separating metadata from data:} the consensus layer in Fabric receives whole transactions as input, but only the transaction IDs are required to decide the transaction order.  We redesign Fabric's transaction ordering service to work with only the transaction IDs, resulting in greatly increased throughput.
    \item \emph{Parallelism and caching:} some aspects of transaction validation can be parallelized while others can benefit from caching transaction data.  We redesign Fabric's transaction validation service by aggressively caching unmarshaled blocks at the comitters and by parallelizing as many validation steps as possible, including endorsement policy validation and syntactic verification.
    \item \emph{Exploiting the memory hierarchy for fast data access on the critical path}: Fabric's key-value store that maintains world state can be replaced with light-weight in-memory data structures whose lack of durability guarantees can be compensated by the blockchain itself.  We redesign Fabric's data management layer around a light-weight hash table that provides faster access to the data on the critical transaction-validation path, deferring storage of immutable blocks to a write-optimized storage cluster. 
    \item \emph{Resource separation:} the peer roles of committer and endorser vie for resources. We introduce an architecture that moves these roles to separate hardware.
\end{enumerate}

Importantly, our optimizations do not violate any APIs or modularity boundaries of Fabric, and therefore they can be incorporated into the planned release of Fabric version 2.0~\cite{FabricRefactor}.  We also outline several directions for future work, which, together with the optimization we propose, have the potential to reach 50,000 transactions per second as required by a credit card company such as Visa~\cite{vukolic2015quest}. 

In the remainder of this paper, Section~\ref{sec:fabric_architecture} gives a brief overview of Hyperledger Fabric, Section~\ref{sec:impr} presents our improved design, Section~\ref{sec:exp} discusses experimental results, Section~\ref{sec:related} places our contributions in the context of prior work, and Section~\ref{sec:conclusions} concludes with directions for future work.


\section{Fabric architecture}
\label{sec:fabric_architecture}

A component of the open-source Hyperledger project hosted by the Linux Foundation, \emph{Fabric} is one of the most actively developed permissioned blockchain systems~\cite{cachin2016architecture}. 
Since Androulaki \textit{et al}~\cite{Androulaki2018a} 
describe the transaction flow in detail, we present only a short synopsis,
focusing on those parts of the system where we propose improvements in Section~\ref{sec:impr}.

To avoid pitfalls with smart-contract determinism and to allow plug-and-play replacement of system components, Fabric is structured differently than other common blockchain systems. Transactions follow an \textit{execute-order-commit} flow pattern instead of the common \textit{order-execute-commit} pattern. Client transactions are first executed in a sandbox to determine their \textit{read-write sets}, i.e., the set of keys read by and written to by the transaction. Transactions are then ordered by an ordering service, and finally validated and committed to the blockchain. This workflow is implemented by nodes that are assigned specific roles, as discussed next. 

\subsection{Node types}
Clients originate transactions, i.e., reads and writes to the blockchain, that are sent to 
Fabric \textit{nodes}\footnote{Because of its permissioned nature, 
all nodes must be known and registered with a membership service provider (MSP), 
otherwise they will be ignored by other nodes.}.
Nodes are either \emph{peers} or \emph{orderers}; some peers are also \textit{endorsers}.
All peers commit blocks to a local copy of the blockchain and 
apply the corresponding changes to a \textit{state database}
that maintains a snapshot of the current \textit{world state}.
Endorser peers are permitted to certify that a transaction is valid  
according to business rules captured in chaincode, Fabric's version of smart contracts. 
Orderers are responsible solely for deciding transaction order, not correctness or validity.
    
\subsection{Transaction flow}
A client sends its transaction to some number of endorsers. Each endorser executes the transaction in a sandbox and computes the corresponding read-write set along with the version number of each key that was accessed. Each endorser also uses business rules to validate the
correctness of the transaction.
The client waits for a sufficient number of endorsements and 
then sends these responses to the orderers, which implement the ordering service. 
The orderers first come to a consensus about the order of incoming transactions and then 
segment the message queue into blocks.
Blocks are delivered to peers, 
who then validate and commit them.
    
\subsection{Implementation details}
To set the stage for a discussion of the improvements in Section~\ref{sec:impr},
we now take a closer look at the orderer and peer architecture.

\subsubsection{Orderer}
After receiving responses from endorsing peers, a client
creates a \emph{transaction proposal} containing a header and a payload. 
The header includes the Transaction ID and Channel ID\footnote{Fabric is
virtualized into multiple \textit{channels}, identified by 
the channel ID.}.
The payload includes the read-write sets and the corresponding version numbers, and endorsing peers' signatures.  The transaction proposal is signed using the client's credentials and sent to the ordering service. 

The two goals of the ordering service are 
(a) to achieve consensus on the transaction order and
(b) to deliver blocks containing ordered transactions to the committer peers. 
Fabric currently uses Apache Kafka, which is based on ZooKeeper~\cite{kafka}, for achieving crash-fault-tolerant consensus. 

When an orderer receives a transaction proposal, it checks if the client is authorized to submit the transaction.
If so, the orderer publishes the transaction proposal to a Kafka cluster, where
each Fabric channel is mapped to a Kafka topic to create a corresponding
immutable serial order of transactions. 
Each orderer then assembles the transactions received from Kafka into blocks,
based either on the maximum number of transactions allowed per block or a block timeout period. 
Blocks are signed using the orderer's credentials and delivered to peers using gRPC~\cite{grpc}. 
        
\subsubsection{Peer}
\label{sec:peer}
On receiving a message from the ordering service 
a peer first unmarshals the header and metadata of the block 
and checks its syntactic structure. 
It then verifies that the signatures of the orderers that 
created this block conform to the specified policy. 
A block that fails any of these tests is immediately discarded. 

After this initial verification, the block is pushed into a queue, guaranteeing 
its addition to the blockchain. However, before that happens, blocks go sequentially through two validation steps and a final commit step. 

During the first validation step, all transactions in the block are unpacked, 
their syntax is checked and their endorsements are validated. 
Transactions that fail this test are flagged as invalid, but are left in the block. 
At this point, only transactions that were created in good faith are still valid. 

In the second validation step, the peer ensures that the interplay between 
valid transactions does not result in an invalid world state. 
Recall that every transaction carries a set of keys 
it needs to read from the world state database (its read set) and a set of keys 
and values it will write to the database (its write set), along with their version numbers recorded by the endorsers. 
During the second validation step, every key in a transaction's read and write sets must still have the same version number.
A write to that key from any prior transaction updates the version number and 
invalidates the transaction. This prevents double-spending. 

In the last step, the peer writes the block,
which now includes validation flags for its transactions,
to the file system.
The keys and their values, i.e., the world state, are 
persisted in either LevelDB or CouchDB, depending on the configuration of the application. 
Moreover, indices to each block and its transactions are stored
in LevelDB to speed up data access.

\section{Design}
\label{sec:impr}
This section presents our changes to the
architecture and implementation of Fabric version 1.2.
This version was released in July~2018, followed by 
the release of version 1.3 in September~2018 and 1.4 in January~2019. 
However, the changes introduced in the recent releases do not interfere with our 
proposal, so we foresee no difficulties in integrating our work with newer versions.
Importantly, our improvements leave the interfaces 
and responsibilities of the individual modules intact, meaning that our changes
are compatible with existing peer or ordering service implementations. 
Furthermore, our improvements are mutually orthogonal and 
hence can be implemented individually. For both orderers and peers, we describe our proposals in ascending order from smallest to largest performance impact compared to their respective changes to Fabric.

\subsection{Preliminaries}
Using a Byzantine-Fault-Tolerant (BFT) consensus algorithm is a critical performance
bottleneck in HyperLedger~\cite{vukolic2015quest}. 
This is because BFT consensus algorithms do not scale 
well with the number of participants. 
In our work, 
we chose to look beyond this obvious bottleneck for three reasons:
\begin{itemize}
\item Arguably, 
    the use of BFT protocols in permissioned blockchains is not as important 
    as in permissionless systems because all participants are known and incentivized to keep the system running in an honest manner.
    \item BFT consensus is being extensively studied~\cite{bano2017consensus} 
    and we expect higher-throughput solutions to emerge in the next year or two.
    \item In practice, Fabric~1.2 does not use a BFT consensus protocol,
    but relies, instead, on Kafka for transaction ordering, as discussed earlier. 
\end{itemize}
For these reasons, 
the goal of our work is not to improve orderer performance using better
BFT consensus algorithms, but to
mitigate 
new issues that arise when the consensus is no 
longer the bottleneck. We first present two improvement to the ordering service,
then a series of improvements to peers.


    \subsection{Orderer improvement I: Separate transaction header from payload}
    \label{sec:orderer1}
In Fabric 1.2, orderers using Apache Kafka send the entire transaction to Kafka for ordering.
Transactions can be several kilobytes in length, resulting in high communication overhead which impacts overall performance.
However, obtaining consensus on the transaction order only
requires transaction IDs, so we can obtain a significant improvement in orderer throughput by sending only transaction IDs
to the Kafka cluster. 

Specifically, on receiving a transaction from a client,
our orderer extracts the transaction ID from the header and publishes this ID to the 
Kafka cluster. The corresponding payload is stored separately in a local data structure by the orderer and
the transaction is reassembled 
when the ID is received back from Kafka. 
Subsequently, as in Fabric, the orderer segments sets of transactions into blocks and delivers
them to peers. Notably, our approach works with any consensus implementation and does not require any modification to the existing ordering interface, allowing us to leverage existing Fabric clients and peer code.

    \subsection{Orderer improvement II: Message pipelining}
        \label{sec:orderer2}

In Fabric~1.2, the ordering service handles incoming transactions from any given client one by one. When a transaction arrives, its corresponding channel is identified, its validity checked against a set of rules and finally it is forwarded to the consensus system, e.g. Kafka; only then can the next transaction be processed. Instead, we implement a pipelined mechanism that can process multiple incoming transactions concurrently, even if they originated from the same client using the same gRPC connection. To do so, we maintain a pool of threads that process incoming requests in parallel, with one thread per incoming request. A thread calls the Kafka API to publish the transaction ID and sends a response to the client when successful. The remainder of the processing done by an orderer is identical to Fabric~1.2.

\begin{figure}[t]
    \centering
    \resizebox{0.7\columnwidth}{!}{\tikzset{every picture/.style={line width=0.5pt}} 

\begin{tikzpicture}[>=stealth, x=0.65pt,y=0.65pt,yscale=-1,xscale=1]

	\pgfdeclarelayer{background}
	\pgfdeclarelayer{foreground}
	\pgfsetlayers{background,main,foreground}


\draw  [fill=white,rounded corners=0.2cm,fill={rgb, 255:red, 255; green, 255; blue, 255 }  ,fill opacity=1, xshift=-6, yshift =6 ] (150,38) -- (220,38) -- (220,78) -- (150,78) -- cycle ;
\draw  [rounded corners=0.2cm,fill={rgb, 255:red, 255; green, 255; blue, 255 }  ,fill opacity=1 , xshift=-3, yshift =3] (142,47) -- (212,47) -- (212,87) -- (142,87) -- cycle ;
\draw  [rounded corners=0.2cm,fill={rgb, 255:red, 255; green, 255; blue, 255 }  ,fill opacity=1 ] (133,56) -- (203,56) -- (203,96) -- (133,96) -- cycle ;
\draw  [rounded corners=0.2cm,fill={rgb, 255:red, 255; green, 255; blue, 255 }  ,fill opacity=1 , xshift=-6, yshift =6] (329,39) -- (399,39) -- (399,79) -- (329,79) -- cycle ;
\draw  [rounded corners=0.2cm,fill={rgb, 255:red, 255; green, 255; blue, 255 }  ,fill opacity=1 , xshift=-3, yshift =3] (314,48) -- (390.5,48) -- (390.5,88) -- (314,88) -- cycle ;
\draw  [rounded corners=0.2cm,fill={rgb, 255:red, 255; green, 255; blue, 255 }  ,fill opacity=1 ] (302.5,57) -- (382,57) -- (382,97) -- (302.5,97) -- cycle ;
\draw [yshift=3]   (220.5,71.98) -- (296.5,71.02) ;
\draw [shift={(298.5,76)}, rotate = 180] [fill={rgb, 255:red, 0; green, 0; blue, 0 }  ][line width=0.75]  [draw opacity=0] (8.93,-4.29) -- (0,0) -- (8.93,4.29) -- cycle    ;
\draw [shift={(218.5,76)}] [fill={rgb, 255:red, 0; green, 0; blue, 0 }  ][line width=0.75]  [draw opacity=0] (8.93,-4.29) -- (0,0) -- (8.93,4.29) -- cycle    ;
\draw    (172.5,100) -- (172.5,130) ;

\draw [shift={(172.5,98)}, rotate = 90] [fill={rgb, 255:red, 0; green, 0; blue, 0 }  ][line width=0.75]  [draw opacity=0] (8.93,-4.29) -- (0,0) -- (8.93,4.29) -- cycle    ;
\draw  [fill={rgb, 255:red, 255; green, 255; blue, 255 }  ,fill opacity=1 , xshift=-2] (158,142) -- (203.5,142) -- (203.5,170) -- (158,170) -- cycle ;
\draw  [fill={rgb, 255:red, 255; green, 255; blue, 255 }  ,fill opacity=1 , xshift=-1] (153,145) -- (198.5,145) -- (198.5,173) -- (153,173) -- cycle ;
\draw  [fill={rgb, 255:red, 255; green, 255; blue, 255 }  ,fill opacity=1 ] (148,148) -- (193.5,148) -- (193.5,176) -- (148,176) -- cycle ;

\draw    (133.5,161) -- (104,161) ;

\draw [shift={(138,161)}, rotate =180] [fill={rgb, 255:red, 0; green, 0; blue, 0 }  ][line width=0.75]  [draw opacity=0] (8.93,-4.29) -- (0,0) -- (8.93,4.29) -- cycle    ;
\draw    (231.5,159) -- (200.5,159) ;

\draw [shift={(233.5,159)}, rotate = 180] [fill={rgb, 255:red, 0; green, 0; blue, 0 }  ][line width=0.75]  [draw opacity=0] (8.93,-4.29) -- (0,0) -- (8.93,4.29) -- cycle    ;
\draw  [fill={rgb, 255:red, 255; green, 255; blue, 255 }  ,fill opacity=1 ] (233,144) -- (310.5,144) -- (310.5,177) -- (233,177) -- cycle ;
\draw    (338.5,160) -- (311.5,160) ;

\draw [shift={(340.5,160)}, rotate = 180] [fill={rgb, 255:red, 0; green, 0; blue, 0 }  ][line width=0.75]  [draw opacity=0] (8.93,-4.29) -- (0,0) -- (8.93,4.29) -- cycle    ;
\draw  [fill={rgb, 255:red, 255; green, 255; blue, 255 }  ,fill opacity=1 ] (340,143) -- (395.5,143) -- (395.5,176) -- (340,176) -- cycle ;
\draw    (264.78,141.98) -- (192.5,99) ;

\draw [shift={(266.5,143)}, rotate = 210.74] [fill={rgb, 255:red, 0; green, 0; blue, 0 }  ][line width=0.75]  [draw opacity=0] (8.93,-4.29) -- (0,0) -- (8.93,4.29) -- cycle    ;

\draw (169,77) node  [align=left] {Kafka };
\draw (343,78) node  [align=left] {Zookeeper};
\draw (286,112) node  [align=left] {orderered TxID};
\draw (149,114) node  [align=left] {TxID};
\draw (171,160) node  [align=left] (txs){Txs};
\draw[->, -latex, xshift=15, yshift=13] (171,160) --node[font=\footnotesize, right, xshift=-2, yshift=-2]{scale} (183,150);
\draw (118,146) node  [align=left] {Tx};
\draw (271,160) node  [align=left](assemble) {Assemble};
\draw (368,160) node  [align=left] (block){Block};

\begin{pgfonlayer}{background}
\node[draw, rounded corners=0.2cm, fit=(txs)(assemble)(block), inner xsep=1em, inner ysep=1.2em](orderer){};
\node[fill=white] at ($(orderer.north)+(90,0)$) {Orderer};
\node[draw, fill=white ,rounded corners=0.2cm, minimum width = 1.6cm, minimum height = 0.93cm] at (69,157) {};
\node[draw, fill=white,rounded corners=0.2cm, minimum width = 1.6cm, minimum height = 0.93cm] at (66,161) {};

\end{pgfonlayer}

\node[draw, fill=white,rounded corners=0.2cm, minimum width = 1.6cm, minimum height = 0.93cm] at (63,165) {Clients};

	\draw[->, -latex] (205,97) --node[font=\footnotesize, right, xshift=-2, yshift=-2]{scale} (215,87);
	
	\draw[->, -latex, xshift=117] (205,97) --node[font=\footnotesize, right, xshift=-2, yshift=-2]{scale} (215,87);

\end{tikzpicture}}
    \caption{New orderer architecture. Incoming transactions are processed concurrently. Their TransactionID is sent to the Kafka cluster for ordering. When receiving ordered TransactionIDs back, the orderer reassembles them with their payload and collects them into blocks.}
    \label{fig:orderer}
\end{figure}
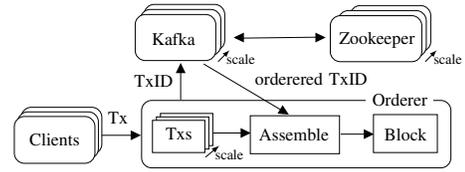

Fig.~\ref{fig:orderer} summarizes the new orderer design, including the separation of transaction IDs from payloads and the scale out due to parallel message processing.

    \subsection{Peer tasks}
Recall from Section \ref{sec:peer} that on receiving a block from an endorser, a Fabric peer carries out the following tasks in sequence:
    \begin{itemize}
        \item Verify legitimacy of the received message
        \item Validate the block header and each endorsement signature for each transaction in the block
        \item Validate read and write sets of the transactions
        \item Update the world state in either LevelDB or CouchDB
        \item Store the blockchain log in the file system, with corresponding indices in LevelDB
    \end{itemize}
Our goal is to maximize transaction throughput on the critical path of the transaction flow. To this end, we performed an extensive call graph analysis to identify performance bottlenecks.

We make the following observations. 
First, the validation of a transaction's read and write set needs fast access to the world state. Thus, we can speed up the process by 
using an in-memory hash table instead of a database (Section \ref{sec:imprp1}). 
Second, the blockchain log is not needed for the transaction flow, so we can defer storing it to a dedicated storage and data analytics server at the end of the transaction flow (Section \ref{sec:imprp2}). 
Third, a peer needs to process new transaction proposals if it is also an endorser. 
However, the committer and endorser roles are distinct, making it possible to dedicate
different physical hardware to each task (Section \ref{sec:imprp4}).
Fourth, incoming blocks and transactions must be validated and resolved at the peer. Most importantly, the validation of the state changes through transaction write sets must be done sequentially,
blocking all other tasks. Thus, it is important to speed up this task as much as possible~(Section~\ref{sec:imprp3}).

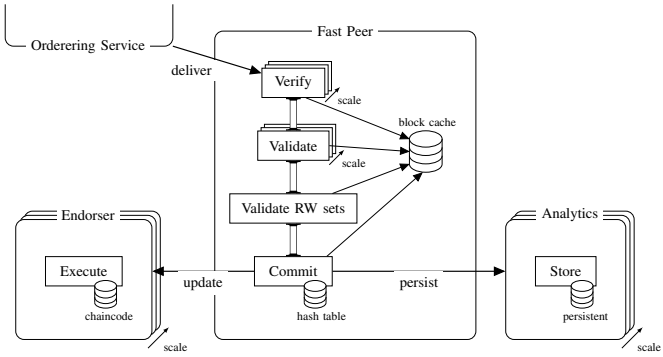
\begin{figure}[t]
        \centering
        \resizebox{\columnwidth}{!}{	\begin{tikzpicture}[>=serif cm]
	\pgfdeclarelayer{background}
	\pgfdeclarelayer{foreground}
	\pgfsetlayers{background,main,foreground}

	\def\layerdiff{0.08cm}
	
	\def\height{0.7cm}
	\def\verwidth{1.5cm}

	\node[draw, fill=white, anchor=west, minimum height = \height, minimum width = \verwidth] at (0,0) (verify1) {};
	\node[draw, fill=white, minimum height = \height, minimum width = \verwidth] at ($(verify1) - (\layerdiff,\layerdiff)$) (verify2) {};
	\node[draw, fill=white, minimum height = \height, minimum width = \verwidth] at ($(verify2) - (\layerdiff,\layerdiff)$) (verify3) {Verify};
	
	\draw[->, -latex] ($(verify3.south east)-(0,0.1)$) --node[font=\footnotesize, right, xshift=0, yshift=-4]{scale} ($(verify1.south east)+(0.2,0.1)$);
	
	\def\valwidth{1.7cm}
	
	\node[draw, fill=white, minimum height = \height, minimum width = \valwidth] at ($(verify1) + (0,-1.5)$) (validate1) {};
	\node[draw, fill=white, minimum height = \height, minimum width = \valwidth] at ($(validate1) - (\layerdiff,\layerdiff)$) (validate2) {};
	\node[draw, fill=white, minimum height = \height, minimum width = \valwidth] at ($(validate2) - (\layerdiff,\layerdiff)$) (validate3) {Validate};
	
	\draw[->, -latex] ($(validate3.south east)-(0,0.1)$) --node[font=\footnotesize, right, xshift=0, yshift=-4]{scale} ($(validate1.south east)+(0.2,0.1)$);
	
	\node[draw, fill=white, minimum height = \height, inner xsep=8] at ($(validate3) + (0,-1.5)$) (valrw) {Validate RW sets};
	
	\def\comwidth{1.3cm}
	\node[draw, fill=white, minimum height = \height, inner xsep=10] at ($(valrw) +(0,-1.5) $) (commit) {Commit};
	
	\path (verify3) edge[double,<->, double distance = 4pt] (validate3);
	\path (validate3) edge[double,<->, double distance = 4pt] (valrw);
	\path (valrw) edge[double,<->, double distance = 4pt] (commit);
	
	\draw[fill=white] ($(commit) +(0.4*\comwidth,-0.5*\height) $) ellipse[x radius = 0.2*\comwidth, y radius = 0.1*\comwidth]
	coordinate (B) at ($(commit) +(0.4*\comwidth,-0.5*\height) - (0.2*\comwidth,0) $) coordinate (A) at ($(commit) +(0.4*\comwidth,-0.5*\height) + (0.2*\comwidth,0) $);	
	\draw[draw] ($(B) +(0,-0.133) $) arc[x radius = 0.2*\comwidth, y radius = 0.12*\comwidth,
	start angle = 180, end angle = 360];	
	\draw[draw] ($(B) +(0,-0.266) $) arc[x radius = 0.2*\comwidth, y radius = 0.12*\comwidth,
	start angle = 180, end angle = 360];		
	\draw ($(B) +(0,-0.4) $) coordinate (C) arc[x radius = 0.2*\comwidth, y radius = 0.1*\comwidth,
	start angle = 180, end angle = 360] coordinate (D);	
	\draw (A) -- (D);
	\draw (B) -- (C);
	\node[anchor = north, font=\footnotesize, yshift=-3, xshift=-3] at ($(D)$) (ht) {hash table};
	
	\draw[fill=white] ($(validate1) +(3cm,0) $) ellipse[x radius = 0.3*\comwidth, y radius = 0.15*\comwidth]
	coordinate (B4) at ($(validate1) +(3cm,0) - (0.3*\comwidth,0) $) coordinate (A4) at ($(validate1) +(3cm,0) + (0.3*\comwidth,0) $);	
	\draw[draw] ($(B4) +(0,-0.2) $) arc[x radius = 0.3*\comwidth, y radius = 0.15*\comwidth,
	start angle = 180, end angle = 360];	
	\draw[draw] ($(B4) +(0,-0.4) $) arc[x radius = 0.3*\comwidth, y radius = 0.15*\comwidth,
	start angle = 180, end angle = 360];		
	\draw ($(B4) +(0,-0.6) $) coordinate (C4) arc[x radius = 0.3*\comwidth, y radius = 0.15*\comwidth,
	start angle = 180, end angle = 360] coordinate (D4);	
	\draw (A4) -- (D4);
	\draw (B4) -- (C4);
	\node[anchor = south, font=\footnotesize, yshift=5, xshift=-10] at ($(A4)$) (cache) {block cache};
	
	\draw[->, -triangle 45] ($(verify3.south) +(0.3,0)$) -- (B4) ;
	
	\draw[->, -triangle 45] (validate3.east) -- ($(B4)-(0,0.3)$);
	\draw[->, -triangle 45] (valrw) -- (C4) ;
	\draw[->, -triangle 45] ($(commit.north) + (0.8,0)$) -- ($(C4)!0.4!(D4) - (0,0.2)$) ;

	\node[draw, fill=white, minimum height = \height, inner xsep=10] at ($(commit) +(6.5,0) $) (store) {Store};	
	
	\draw[fill=white] ($(store) +(0.3*\comwidth,-0.5*\height) $) ellipse[x radius = 0.2*\comwidth, y radius = 0.1*\comwidth]
	coordinate (B2) at ($(store) +(0.3*\comwidth,-0.5*\height) - (0.2*\comwidth,0) $) coordinate (A2) at ($(store) +(0.3*\comwidth,-0.5*\height) + (0.2*\comwidth,0) $);	
	\draw[draw] ($(B2) +(0,-0.133) $) arc[x radius = 0.2*\comwidth, y radius = 0.12*\comwidth,
	start angle = 180, end angle = 360];	
	\draw[draw] ($(B2) +(0,-0.266) $) arc[x radius = 0.2*\comwidth, y radius = 0.12*\comwidth,
	start angle = 180, end angle = 360];	
	\draw ($(B2) +(0,-0.4) $) coordinate (C2) arc[x radius = 0.2*\comwidth, y radius = 0.1*\comwidth,
	start angle = 180, end angle = 360] coordinate (D2);
	\draw (A2) -- (D2);
	\draw (B2) -- (C2);
	\node[anchor = north, font=\footnotesize, yshift=-3, xshift=-5pt] at ($(D2)$) (db) {persistent};

	\node[draw, fill=white, minimum height = \height, inner xsep=10] at ($(commit) -(5,0) $) (execute) {Execute};
	
	\draw[fill=white] ($(execute) +(0.4*\comwidth,-0.5*\height) $) ellipse[x radius = 0.2*\comwidth, y radius = 0.1*\comwidth]
	coordinate (B3) at ($(execute) +(0.4*\comwidth,-0.5*\height) - (0.2*\comwidth,0) $) coordinate (A3) at ($(execute) +(0.4*\comwidth,-0.5*\height) + (0.2*\comwidth,0) $);	
	\draw[draw] ($(B3) +(0,-0.133) $) arc[x radius = 0.2*\comwidth, y radius = 0.12*\comwidth,
	start angle = 180, end angle = 360];	
	\draw[draw] ($(B3) +(0,-0.266) $) arc[x radius = 0.2*\comwidth, y radius = 0.12*\comwidth,
	start angle = 180, end angle = 360];	
	\draw ($(B3) +(0,-0.4) $) coordinate (C3) arc[x radius = 0.2*\comwidth, y radius = 0.1*\comwidth,
	start angle = 180, end angle = 360] coordinate (D3);
	\draw (A3) -- (D3);
	\draw (B3) -- (C3);
	\node[anchor = north, font=\footnotesize, yshift=-3, xshift=-5pt] at ($(D3)$) (chaincode) {chaincode};
	
	\begin{pgfonlayer}{background}
	
		\node[draw, rounded corners=0.2cm, fit=(commit) (valrw)(validate3)(verify3)(D)(cache), inner xsep=1em, inner ysep=2.5em](fastpeer){};
		\node[fill=white] at (fastpeer.north) {Fast Peer};
	
		\node[draw, fill=white, rounded corners=0.2cm, fit=(execute)(D3), inner xsep=2em, inner ysep=2.5em, xshift=0.2cm, yshift=0.2cm](endorser3){};
		\node[draw, fill=white, rounded corners=0.2cm, fit=(execute)(D3), inner xsep=2em, inner ysep=2.5em, xshift=0.1cm, yshift=0.1cm](endorser2){};
		\node[draw, fill=white, rounded corners=0.2cm, fit=(execute)(D3), inner xsep=2em, inner ysep=2.5em](endorser){};
		\node[fill=white] at (endorser2.north) {Endorser};
		\draw[->, -latex] ($(endorser.south east)-(0.1,0.2)$) --node[font=\footnotesize, below right]{scale} ($(endorser3.south east)+(0.2,0.1)$);

		\node[draw, fill=white, rounded corners=0.2cm, fit=(store) (D2), inner xsep=2em, inner ysep=2.5em, xshift=0.2cm, yshift=0.2cm](persist3){};
		\node[draw, fill=white, rounded corners=0.2cm, fit=(store) (D2), inner xsep=2em, inner ysep=2.5em, xshift=0.1cm, yshift=0.1cm](persist2){};
		\node[draw, fill=white, rounded corners=0.2cm, fit=(store) (D2), inner xsep=2em, inner ysep=2.5em](persist){};
		\node[fill=white] at (persist2.north) {Analytics};
		\draw[->, -latex] ($(persist.south east)-(0.1,0.2)$) --node[font=\footnotesize, below right]{scale} ($(persist3.south east)+(0.2,0.1)$);

	\end{pgfonlayer}
	
	\draw[thick,->, -triangle 45] (commit.west) -- node[below, fill=white] {update} (commit.west -| endorser.east) ;
	
	\draw[thick,->, -triangle 45] (commit.east) -- node[below, fill=white]{persist} (commit.east -| persist.west);
	
	\node[] at ($(verify3) + (-3,1)$) (deliver){};

	\draw[thick,->, -triangle 45] (deliver.south east) -- node[below left, fill=white]{deliver}(verify3);	
	
	\draw[rounded corners=0.2cm] ($(deliver.south east) +(-4,1)$) |- (deliver.south east) -- ($(deliver.south east) +(0,1)$) ;	
	
	\node[fill=white, xshift=-1.9cm] at (deliver.south) {Orderering Service};

	\end{tikzpicture}}
        \caption{New peer architecture. The fast peer uses an in-memory hash table to store the world state. The validation pipeline is completely concurrent, validating multiple blocks and their transactions in parallel. The endorser role and the persistent storage are separated into scalable clusters and given validated blocks by the fast peer. All parts of the pipeline make use of unmarshaled blocks in a cache.}
        \label{fig:peer_newarch}
    \end{figure}

Finally, significant performance gains can be obtained by caching the results of the \emph{Protocol Buffers}~\cite{GoogleDevelopers2018} unmarshaling of blocks (Section \ref{sec:imprp5}).
We detail this architectural redesign, including the other proposed peer improvements, in Figure~\ref{fig:peer_newarch}.
    
\subsection{Peer improvement I: Replacing the world state database with a hash table}
    \label{sec:imprp1}
    The world state database must be looked up and updated \textit{sequentially} for each
    transaction to guarantee consistency across all peers. 
    Thus, it is critical that updates to this data store happen at the highest possible transaction rate. 
    
    We believe that for common scenarios, such as for tracking of wallets or assets on the ledger,
    the world state is likely to be relatively small. Even if billions of keys need to be stored, 
    most servers can easily keep them in memory. Therefore, we propose
    using an in-memory hash table, instead of LevelDB/CouchDB, to store world state. 
    This eliminates hard drive access when updating the world state. It also eliminates costly database system guarantees (i.e., ACID properties)
    that are unnecessary due to redundancy guarantees of the blockchain itself, further boosting the performance. 
    Naturally, such a replacement is susceptible to node failures due to the use of volatile memory, so 
    the in-memory hash table must be augmented by stable storage. 
    We address this issue in Section~\ref{sec:imprp2}.
    
    
\subsection{Peer improvement II: Store blocks using a peer cluster}
    \label{sec:imprp2}

       By definition, blocks are immutable. This makes them ideally suited to append-only data stores. 
       By decoupling data storage from the remainder of a peer's tasks, 
       we can envision many types of data stores for blocks and world state backups, including
       a single server storing blocks and world state backups in its file system, as Fabric does currently;
       a database or a key-value store such as LevelDB or CouchDB.
       For maximum scaling, we propose the use of a distributed storage cluster. Note that with this solution, each storage server contains only a fraction of the chain, motivating the use of distributed data processing tools such as Hadoop MapReduce or Spark\footnote{However, our current implementation does not include such a storage system.}. 
       
\subsection{Peer improvement III: Separate commitment and endorsement}
    \label{sec:imprp4}
    
In Fabric~1.2, endorser peers are also responsible for committing blocks. Endorsement is an expensive operation, as is commitment.
While concurrent transaction processing on a cluster of endorsers could potentially improve application performance, the additional work to replicate commitments on every new node effectively nullifies the benefits.
Therefore, we propose to split these roles.
    
    Specifically, in our design, a committer peer executes the validation pipeline and then 
    sends validated blocks to a cluster of endorsers who only apply the changes to their world state 
    without further validation. This step allows us to free up resources on the peer. 
    Note that such an endorser cluster, which can scale out to meet demand, only splits off the endorsement role of a peer to dedicated hardware. Servers in this cluster  are not equivalent to full fledged endorsement peers in Fabric~1.2.

\subsection{Peer improvement IV: Parallelize validation}
    \label{sec:imprp3}
    
  Both block and transaction header validation,
  which include checking permissions of the sender, enforcing endorsement policies and syntactic verification, 
  are highly parallelizable. We extend the concurrency efforts of Fabric~1.2 by introducing a complete validation pipeline. 
  
  Specifically, for each incoming block, one go-routine is allocated to shepherd it through the block validation phase.
  Subsequently, each of these go-routines makes use of the go-routine pool that already exists in Fabric~1.2 for transaction validation. Therefore, at any given time, multiple blocks and their transactions are checked for validity in parallel. Finally, all read-write sets are validated sequentially by a single go-routine in the correct order.
  This enables us to utilize the full potential of multi-core server CPUs.

\subsection{Peer improvement V: Cache unmarshaled blocks}
    \label{sec:imprp5}
    
Fabric uses gRPC for communication between nodes in the network. 
To prepare data for transmission, \textit{Protocol Buffers} are used for serialization. 
To be able to deal with application and software upgrades over time, 
Fabric's block structure is highly layered, 
where each layer is marshaled and unmarshaled separately. 
This leads to a vast amount of memory allocated to convert
byte arrays into data structures. 
Moreover, Fabric~1.2 does not store previously unmarshaled data in a cache,
so this work has to be redone whenever the data is needed. 

To mitigate this problem, we propose a temporary cache of unmarshaled data. 
Blocks are stored in the cache while in the validation pipeline and 
retrieved by block number whenever needed. 
Once any part of the block becomes unmarshaled, it is stored with the block for reuse. 
We implement this as a cyclic buffer that is as large as the validation pipeline. Whenever a block is committed, a new block can be admitted to the pipeline and automatically overwrites the existing cache location of the committed block. 
As the cache is not needed after commitment and it is guaranteed that a new block only arrives after an old block leaves the pipeline, this is a safe operation. Note that unmarshaling only adds data to the cache, it never mutates it. 
Therefore, lock-free access can be given to all go-routines in the validation pipeline. In a worst-case scenario, multiple go-routines try to access the same (but not yet unmarshaled) data and all proceed to execute the unmarshaling in parallel. 
Then the last write to the cache wins, 
which is not a problem because the result would be the same in either case.

Call graph analysis shows that, even with these optimizations, memory (de)allocation due to 
unmarshaling is still responsible for the biggest share of the overall execution time. This
is followed by gRPC call management and cryptographic computations. The last two areas, however, are beyond the scope of this work.

\section{Results}
    \label{sec:exp}
This section presents an experimental performance evaluation of our architectural improvements.
We used fifteen local servers connected by a 1~Gbit/s switch. 
Each server is equipped with two Intel\textsuperscript{\tiny\textregistered}~Xeon\textsuperscript{\tiny\textregistered}~CPU~E5-2620~v2 processors at 2.10~GHz, for a total of 24 hardware threads and 64~GB of RAM. 
We use Fabric~1.2 as the base case and add our improvements step by step for comparison.
By default, Fabric is configured to use LevelDB as the peer state database and 
the orderer stores completed blocks in-memory, rather than on disk. 
Furthermore, we run the entire system without using docker containers to avoid additional overhead.
    
    While we ensured that our implementation did not change the validation behaviour of Fabric, all tests were run with non-conflicting and valid transactions. This is because valid transactions must go through every validation check and their write sets will be applied to the state database during commitment.
    In contrast, invalid transactions can be dropped. Thus, our results evaluate the worst case performance.
    
   For our experiments which focus specifically on either the orderer or the committer, we isolate the respective system part. In the orderer experiments, we send pre-loaded endorsed transactions from a client to the orderer and have a mock committer simply discard created blocks. Similarly, during the benchmarks for the committer, we send pre-loaded blocks to the committer and create mocks for endorsers and the block store which discard validated blocks.
   
   Then, for the end-to-end setup, we implement the full system: Endorsers endorse transaction proposals from a client based on the replicated world state from validated blocks of the committer; the orderer creates blocks from endorsed transactions and sends them to the committer; the committer validates and commits changes to its in-memory world state and sends validated blocks to the endorsers and the block storage; the block storage uses the Fabric~1.2 data management to store blocks in its file system and the state in LevelDB. We do not, however, implement a distributed 
   block store for scalable analytics; that is beyond the scope of this work.
    
    For a fair comparison,
    we used the same transaction chaincode for all experiments:
    Each transaction simulates a money transfer from one account to another, reading and making changes to two keys in the state database. These transactions carry a payload of 2.9~KB, which is typical~\cite{sousa2018byzantine}. Furthermore, we use the default endorsement policy of accepting a single endorser signature.

    \subsection{Block transfer via gRPC}
    \label{sec:grpc}
    
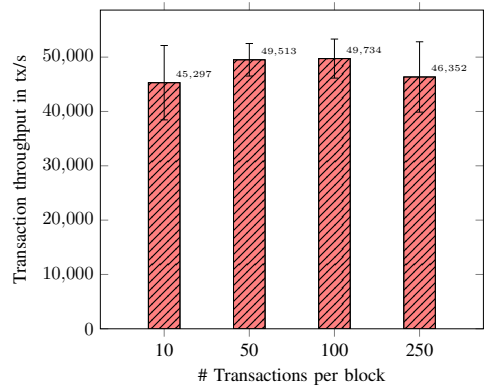
\begin{figure}[t]
        \centering
        \begin{tikzpicture}[scale = 0.7]
        \begin{axis}[
        ybar,
        symbolic x coords={10, 50,100,250},
        scaled ticks=false, 
        tick label style={/pgf/number format/fixed},
        legend style={at={(1,1.03)}, anchor=north east, font=\small},
        legend cell align=left,
        enlarge x limits=0.25,
        ymin=0,
        width= \linewidth,
        bar width=0.6cm,
        ylabel= Transaction throughput in tx/s,
        xlabel = {\# Transactions per block},
        nodes near coords,
        every node near coord/.append style={
            font=\tiny,
            /pgf/number format/precision=0,
            yshift=10, xshift=16, anchor=north}
        ]       
        
        \addplot[fill = red!50, postaction={pattern=north east lines},error bars/.cd, y dir = both,y explicit] coordinates {
        (10,45296.661633)+-(7875.252206,6843.464971)
(50,49512.556706)+-(4364.061518,2994.326095)
(100,49733.911785)+-(9737.804632,3592.449817)
(250,46351.842433)+-(7431.284192,6474.521368)};
        \end{axis}
        \end{tikzpicture}
        \caption{Throughput via gRPC for different block sizes.
        }
        \label{fig:grpc}
\end{figure}

    We start by benchmarking the gRPC performance.  
    We pre-created valid blocks with different numbers of transactions in them, 
    sent them through the Fabric gRPC interface 
    from an orderer to a peer, and then immediately discarded them. The results of this experiment are shown in Figure~\ref{fig:grpc}. 
    
    We find that for block sizes from 10 to 250 transactions,
    which are the sizes that lead to the best performance in the following sections,
    a transaction throughput rate of more that 40,000 transactions/s is sustainable.
    Comparing this with the results from our end-to-end tests in Section~\ref{sec:end2end}, 
    it is clear that in our environment, network bandwidth and the 1~Gbit/s switch used in the server rack are not the performance bottlenecks.
    
        \begin{figure}[t]
    \centering
    \begin{tikzpicture}[scale = 0.7]
        \begin{axis}[
        ybar,
        symbolic x coords={0,512,1024,2048,4096},
        scaled ticks=false, 
        tick label style={/pgf/number format/fixed},
        legend style={at={(1,1.03)}, anchor=north east, font=\small},
        legend cell align=left,
	    enlarge x limits=0.17,
        ymin=0,
        width= \linewidth,
        bar width=0.3cm,
        ylabel= Transaction throughput in txs/s,
        xlabel = {Payload size in Bytes},
        nodes near coords,
        every node near coord/.append style={
            font=\tiny,
            /pgf/number format/precision=0,
            yshift=15,xshift=-5, anchor=south}
        ]       
        
        \addplot[fill = red!50, postaction={pattern=north east lines},error bars/.cd, y dir = both,y explicit] coordinates {
        (0,14940)+-(2440,1726)
(512,8619)+-(1477,1380)
(1024,7328)+-(1078,364)
(2048,6215)+-(659,927)
(4096,4362)+-(362,399)};
        \addplot[fill = blue!50, postaction={pattern=north west lines},error bars/.cd, y dir = both,y explicit] coordinates {
        (0,16857)+-(2571,3142)
(512,14980)+-(3869,1686)
(1024,16928)+-(4428,3071)
(2048,16190)+-(6190,3809)
(4096,12400)+-(1289,1884)};
        \addplot[fill = black!25, postaction={pattern=crosshatch dots},error bars/.cd, y dir = both,y explicit] coordinates {
        (0,28534)+-(5762,5444)
(512,25249)+-(3192,3223)
(1024,24289)+-(4184,3975)
(2048,21719)+-(2405,3322)
(4096,17833)+-(861,500)};
        \legend{Fabric 1.2, Opt O-I, Opt O-II}
        \end{axis}
        \end{tikzpicture}
    \caption{Effect of payload size on orderer throughput.}
    \label{fig:broadcast}
\end{figure}
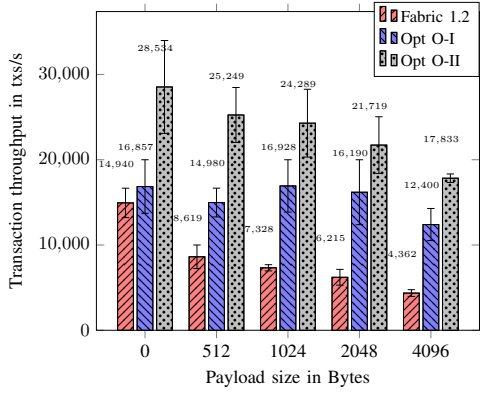

    \subsection{Orderer throughput as a function of message size}

In this experiment, we set up multiple clients that send transactions 
to the orderer and monitor the time it takes to send 100,000 transactions. 
We evaluate the rate at which an orderer can order transactions in Fabric 1.2 and compare it to our improvements: 
\begin{itemize}
    \item \textbf{Opt O-I}: only Transaction ID is published to Kafka (Section \ref{sec:orderer1})
    \item \textbf{Opt O-II}: parallelized incoming transaction proposals from clients (Section~\ref{sec:orderer2})
\end{itemize}

Figure~\ref{fig:broadcast} shows the transaction throughput for different payload sizes. 
In Fabric 1.2, transaction throughput decreases as payload size increases due to the overhead of sending large messages to Kafka. 
However, when we send only the transaction ID to Kafka (Opt O-1), we can almost triple the average throughput (2.8$\times$) for a payload size of 4096~KB.  
Adding optimization O-2 leads to an average throughput of 4$\times$ over the base Fabric 1.2. In particular,
for the 2~KB payload size, we increase orderer performance from 6,215~transactions/s to 21,719~transactions/s, a 
ratio of nearly 3.5x.

    \subsection{Peer experiments}
    \label{sec:peerex}
    In this section, we describe tests on a single peer in isolation 
    (we present the results of an end-to-end evaluation in Section~\ref{sec:end2end}).
    Here, we pre-computed blocks and sent them to a peer as we did in the gRPC experiments 
    in Section~\ref{sec:grpc}. The peer then 
    completely validates and commits the blocks. 
    
    The three configurations shown in the figures compared to Fabric~1.2 \textit{cumulatively} incorporate our improvements (i.e., Opt P-II incorporate Opt P-I, and Opt P-III incorporates both prior improvements):
    \begin{itemize}
        \item \textbf{Opt P-I} LevelDB replaced by an in-memory hash table
        \item \textbf{Opt P-II} Validation and commitment completely parallelized; block storage and endorsement offloaded to a separate storage server via remote gRPC call
        \item \textbf{Opt P-III} All unmarshaled data cached and accessible to the entire validation/commitment pipeline.
    \end{itemize}
    \subsubsection{Experiments with fixed block sizes}

\begin{figure}[t]
            \centering
            \begin{tikzpicture}[scale = 0.7]
		\begin{axis}[
		ybar,
		scaled ticks=false, 
		tick label style={/pgf/number format/fixed},
		legend style={at={(1,1.03)}, anchor=north east, font=\small},
		legend cell align=left,
		width= \linewidth,
		enlarge x limits=0.25,
		xtick=data,
		bar width=0.6cm,
		symbolic x coords = {Fabric 1.2, Opt P-I, Opt P-II, Opt P-III},
		ymin=0, 
		ylabel= Transaction throughput in txs/s,
		nodes near coords,
		every node near coord/.append style={
			font=\tiny,
			/pgf/number format/precision=0,
			yshift=16, xshift=0, anchor=north}
		]		
		\addplot[fill = red!50, postaction={pattern=north east lines},error bars/.cd, y dir = both,y explicit] coordinates {
		(Fabric 1.2,3185.6802345445667)+-(0,62.9390956770419)
		(Opt P-I,7551.183558697136)+-(0,505.0371437639384) 
		(Opt P-II,9117.132420110063)+-(0,829.8109346816813)
		(Opt P-III,21252.348814708508)+-(0,225.62615463189397)};
		\end{axis}
		\end{tikzpicture}
            \caption{Impact of our optimizations on peer throughput.
            }
            \label{fig:peertp}
        \end{figure}
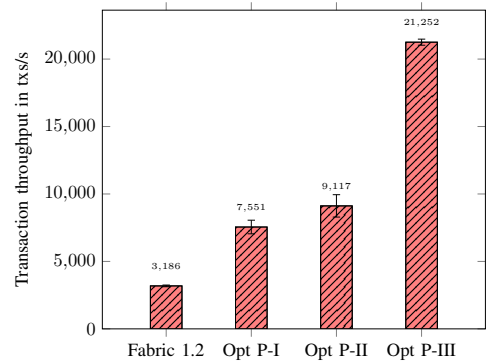

     Fig.~\ref{fig:peerlat} and~\ref{fig:peertp} show the results 
     from validation and commitment of 100,000 transactions for a single run, repeated 1000 times. 
     Transactions were collected into blocks of 100 transactions\footnote{We experimentally determined that peer throughput was maximized at this block size.}. We first discuss latency, then throughput.
     
      \begin{figure}[t]
         \hspace{1.5cm}
        \raisebox{0.4cm}{\begin{tikzpicture}[scale = 0.7]
		\begin{axis}[
		ybar,
		scaled ticks=false, 
		tick label style={/pgf/number format/fixed},
		legend style={at={(1,1.03)}, anchor=north east, font=\small},
		legend cell align=left,
		width= \linewidth,
		enlarge x limits=0.25,
		xtick=data,
		bar width=0.6cm,
		symbolic x coords = {Fabric 1.2, Opt P-I, Opt P-II, Opt P-III},
		ymin=0, 
		ylabel= Block latency in ms,
		nodes near coords,
		every node near coord/.append style={
			font=\tiny,
			/pgf/number format/precision=2,yshift=20, xshift=0}
		]		
		\addplot[fill = red!50, postaction={pattern=north east lines},error bars/.cd, y dir = both,y explicit] coordinates {
		(Fabric 1.2,0036.22903930330328)+-(0,0005.507771872910765)
		(Opt P-I,0020.25933015963855)+-(0,0004.6960498203214825) 
		(Opt P-II,0022.377281245967755)+-(0,0004.595815261553954)
		(Opt P-III,0012.362696652257421)+-(0,0002.2453602701032124)};
		\end{axis}
		\end{tikzpicture}}
        \caption{Impact of our optimizations on peer block latency. 
        }
        \label{fig:peerlat}
    \end{figure}
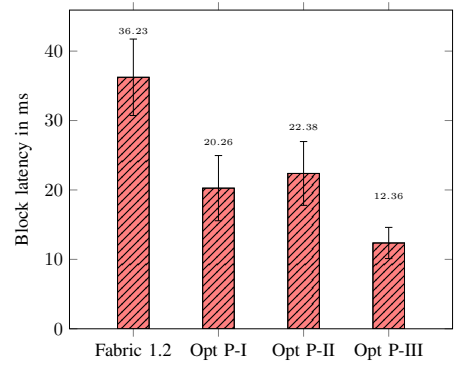
    
        Because of batching, we show the latency per block, rather than per-transaction latency. 
        The results are in line with our self-imposed goal of introducing no additional latency as
        a result of
    increasing throughput; in fact our performance improvements decrease peer latency to a third of its original value (note that these experiments do not take network delay into account). Although the pipelining introduced in Opt~P-II 
        generates some additional latency, the other optimizations more than compensate for it.
        
        By using a hash table for state storage (Opt P-I), we are able to more than double the throughput of a Fabric~1.2 peer from about 3200 to more than 7500 transactions/s.
        Parallelizing validation (Opt P-II) 
        adds an improvement of roughly 2,000 transactions per second. 
        This is because, as Figure~\ref{fig:peer_newarch} shows, only the first two validation steps can be parallelized and scaled out. Thus, the whole pipeline performance is governed by the throughput of read and write set validation and commitment. 
        Although commitment is almost free when using Opt P-I,  it is not until the introduction of the unmarshaling cache in Opt P-III that Opt P-II pays off. The cache drastically reduces the amount of work for the CPU, freeing up resources to validate additional blocks in parallel. 
        With all peer optimizations taken together, we increase a peer's commit
        performance by 7x from about 3200 transactions/s to over 21,000 transactions/s.

        \subsubsection{Parameter sensitivity}
        
        As discussed in Section \ref{sec:peerex}, parallelizing block and transaction validation 
        at the peer is critical. However, it is not clear how much parallelism is necessary
        to maximize performance. Hence,
        we explore the degree to which a peer's performance can be tuned by varying two parameters:
        \begin{itemize}
            \item The number of go-routines concurrently shepherding blocks in the validation pipeline
            \item The number of go-routines concurrently  validating transactions
        \end{itemize}
        
       We controlled the number of active go-routines in the system using semaphores, 
        while allowing multiple blocks to concurrently enter the validation pipeline.
        This allows us to control the level of parallelism in block header validation 
        and transaction validation through two separate go-routine pools. 
        
\begin{figure}[t]
    \centering
    \includegraphics[width=0.722\columnwidth]{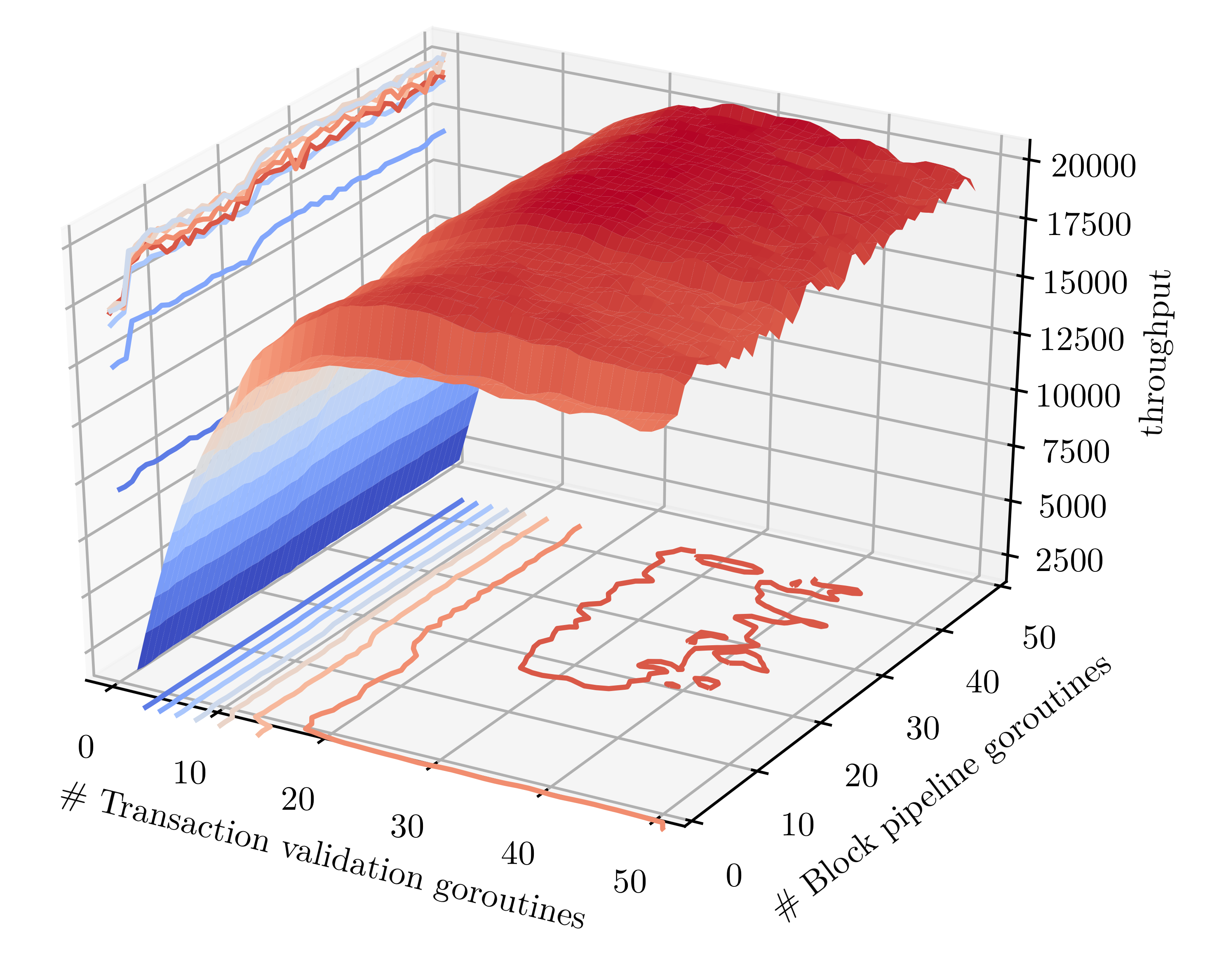}
    \caption{Parameter sensitivity study for blocks containing 100 transactions and a server with 24 CPU cores. We scale the number of blocks that are validated in parallel and the number of transactions per block that are validated in parallel independently.}
    \label{fig:pipe}
\end{figure}

        For a block size of 100 transactions, 
        Figure~\ref{fig:pipe} shows the throughput when varying the number of go-routines. 
        The total number of threads in the validation pipeline 
        is given by the sum of the two independent axes.
        For example, we achieve maximum throughput for 25 transaction validation go-routines and 31 concurrent blocks in the pipeline, totalling 56 go-routines for the pipeline. 
        While we see a small performance degradation through thread management overhead when there are too many threads, the penalty for starving the CPU with too few parallel executions is drastic. Therefore, we suggest as a default that there be at least twice as many go-routines as there are physical threads in a given machine. 
        
       We now investigate the dependence of peer throughput on block size. 
       Each block size experiment was performed with the best tuned go-routine parameters from the previous test.
       All configurations used around $24 \pm 2$ transaction validation go-routines and 
       $30 \pm 3$ blocks in the pipeline. Again, we split 100,000 transactions among blocks of a given size for a single benchmark run and repeated the experiment 1,000 times. We chose to scan the block size space on a logarithmic scale to get an overview of a wide spectrum. 
       
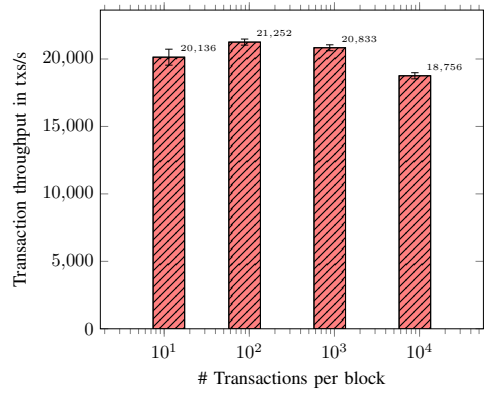
\begin{figure}
            \centering
            \begin{tikzpicture}[scale = 0.7]
		\begin{semilogxaxis}[
		ybar=-12,
		scaled ticks=false, 
		tick label style={/pgf/number format/fixed},
		legend style={at={(1,1.03)}, anchor=north east, font=\small},
		legend cell align=left,
		enlarge x limits=0.25,
		ymin=0,
		width= \linewidth,
		bar width=0.6cm,
		ylabel= Transaction throughput in txs/s,
		xlabel= \# Transactions per block,
		nodes near coords,
		every node near coord/.append style={
			font=\tiny,
			/pgf/number format/precision=0,
			yshift=10, xshift=16, anchor=north}
		]		
		
		\addplot[fill = red!50, postaction={pattern=north east lines},error bars/.cd, y dir = both,y explicit] coordinates {
		(100,21252.348814708508)+-(0,225.62615463189397) 
		(1000,20833.366045149098)+-(0,218.77948039784587)
		(10000,18755.58638880042)+-(0,227.3248586508303)};
		\addplot[fill = red!50, postaction={pattern=north east lines},error bars/.cd, y dir = both,y explicit] coordinates {
		(10,20135.508228514387)+-(0,591.8579346525523)};
		\end{semilogxaxis}
		\end{tikzpicture}
            \caption{Throughput dependence on block size for optimally tuned configuration}
                        \label{fig:blocksize}
        \end{figure}

       The results are shown in Figure~\ref{fig:blocksize}. We find that a block size of 100 
       transactions/block gives the best throughput with just over 21,000 transactions per second.
       We also investigated small deviations from this block size.
       We found that performance differences for block sizes between 50 and 500 were very minor, 
       so we opted to fix the block size to 100 transactions.


        \subsection{End-to-end throughput}
        \label{sec:end2end}
        
        \begin{table}[b]
            \renewcommand{\arraystretch}{1.5}
            \centering
            \caption{End-To-End Throughput}
            \begin{tabular}{|c|r| r|}
               \hline
                &Fabric~1.2 & FastFabric \\
                \hline
                 Transactions/s & $3185 \pm 62$ & $19112 \pm 811$\\
                    \hline
            \end{tabular}
            
            \label{tab:end2end}
        \end{table}
        
        We now discuss the end-to-end throughput achieved by combining all of our optimizations, i.e., Opt. O-II combined with Opt. P-III, compared to our measurements of unmodified Fabric~1.2. 
        
        We set up a single orderer that uses a cluster of three ZooKeeper servers and three Kafka servers, with the default topic replication factor of three, and connect it to a peer. Blocks from this peer are sent to a single data storage server that stores world state in LevelDB and blocks in the file system. For scale-out, five endorsers replicate the peer state and provide sufficient throughput to deal with client
        endorsement load. Finally, a client is installed on its own server; this client requests endorsements from the five endorser servers and sends endorsed transactions to the ordering service. This uses a total of fifteen servers connected to the same 1~Gbit/s switch in our local data center. 
        
        We send a total of 100,000 endorsed transactions from the client to the orderer, which batches them to blocks of size 100 and delivers them to the peer. To estimate throughput,
        we measure the time between committed blocks on the peer and take the mean over a single run. These runs are repeated 100 times. 
        Table~\ref{tab:end2end} shows a significant improvement of $6$-$7 \times$ compared
        to our baseline Fabric~1.2 benchmark.
        
\section{Related Work}
\label{sec:related}

Hyperledger Fabric  is a recent system that is still undergoing rapid development and significant changes in its architecture.
Hence, there is relatively little work on the performance analysis of the system or suggestions for architectural improvements.
 Here, we survey  recent work on techniques to improve the performance of Fabric.
 
 The work closest to ours is by
Thakkar \textit{et al}~\cite{Thakkar2018} who study the impact of various
configuration parameters on the performance of Fabric.  They find that the major bottlenecks are
repeated validation of X.509 certificates during
endorsement policy verification, sequential policy validation of transactions
in a block, and  state validation during the commit phase.
They introduce aggressive caching of verified endorsement
certificates (incorporated into Fabric version 1.1, hence part of our evaluation), 
parallel verification of endorsement
policies, and batched state validation and commitment.
These improvements increase overall throughput by a factor of 16.
We also parallelize verification at the commiters and go one step further in
replacing the state database with a more efficient data structure, a hash table.

In recent work, Sharma \textit{et al}~\cite{sharma2018databasify} study the use of database techniques, i.e., transaction reordering and early abort, to improve the performance of Fabric. Some of their ideas related to early identification of conflicting transactions are orthogonal to ours and can be incorporated into our solution.  However, some ideas, such as having the orderers drop conflicting transactions, are not compatible with our solution. First, we deliberately do not send transaction read-write sets to the orderers, only transaction IDs. Second, we chose to keep to Fabric's design goal of allocating different tasks to different types of nodes, so our orderers do not examine the contents of the read and write sets. An interesting direction for future work would be to compare these two approaches under different transaction workloads to understand when the overhead of sending full transaction details to the orderers is worth the early pruning of conflicting transactions.

 It is well known that Fabric's orderer component can be a performance bottleneck due to the message communication overhead of Byzantine fault tolerant (BFT) consensus protocols.  Therefore, it is important to use an efficient implementation of a BFT  protocol in the orderer.
 Sousa \textit{et al}~\cite{sousa2018byzantine} study the use of the well-known BFT-SMART \cite{sousa2013state} implementation as a part of Fabric and shows that, using this implementation within a single datacenter, a throughput of up to 30,000 transactions/second is achievable.
  However,  unlike our work, the commiter component is not benchmarked  and  the end-to-end  performance is not addressed.
  
  Androulaki \textit{et al}~\cite{androulaki2018channels} study the use of channels for scaling Fabric. However, this work does not
   present a performance evaluation to quantitatively establish the benefits from their approach.
  
  Raman \textit{et al}~\cite{raman2018trusted} study the use of lossy compression to reduce the communication cost of sharing 
  state between Fabric endorsers and validators when a blockchain is used 
  for storing intermediate results arising from the analysis of large datasets. However, their approach is only applicable 
  to scenarios which are
  insensitive to lossy compression, which is not the general case for blockchain-based applications. 
  
  Some studies have examined the performance of Fabric without suggesting internal architectural changes to 
  the underlying system.
 For example, Dinh \textit{et al} use BlockBench~\cite{Dinh2017d}, a tool to study the performance of private blockchains, 
 to study the performance of Fabric, comparing it with that of Ethereum and Parity. They found that the version of Fabric they studied did not
 scale beyond 16 nodes due to congestion in the message channel. 
 Nasir \textit{et al}~\cite{nasir2018performance} compare the performance of Fabric 0.6 and 1.0, finding, unsurprisingly, that 
the 1.0 version outperforms the 0.6 version.
 Baliga \textit{et al}~\cite{Baliga2018} showed that application-level parameters such as the
read-write set size of the transaction and
chaincode and event payload sizes significantly impact transaction latency.
 Similarly, Pongnumkul \textit{et al}~\cite{Pongnumkul2017} compare the performance of Fabric and Ethereum for a 
 cryptocurrency workload, finding that Fabric outperforms Ethereum in all metrics.
 Bergman ~\cite{bergman2018permissioned}
compares the performance of Fabric to Apache Cassandra in similar environments and finds that,
for a small number of peer nodes,
Fabric has a lower latency for linearizable transactions in read-heavy workloads than Cassandra. 
On the other hand, with a larger number of nodes, or write-heavy workloads, Cassandra has better performance.

\section{Conclusions}
\label{sec:conclusions}

The main contribution of this work is to show how a permissioned blockchain framework such as Hyperledger Fabric can be re-engineered to support nearly 20,000 transactions per second, a factor of almost 7 better than prior work.  We accomplished this goal by implementing a series of independent optimizations focusing on I/O, caching, parallelism and efficient data access.  In our design, orderers only receive transaction IDs instead of full transactions, and validation on peers is heavily parallelized. We also use aggressive caching and we leverage light-weight data structures for fast data access on the critical path.
In future work, we hope to further improve the performance of Hyperledger Fabric by:

\begin{itemize}
    \item Incorporating an efficient BFT consensus algorithm such as RCanopus \cite{rcanopus}
    \item Speeding up the extraction of transaction IDs for the orderers without unpacking the entire transaction headers
    \item Replacing the existing cryptographic computation library with a more efficient one
    \item Providing further parallelism by assigning a separate ordering and fast peer server per channel
    \item Implementing an efficient data analytics layer using a distributed framework such as Apache Spark \cite{spark}
\end{itemize}


\bibliographystyle{IEEEtranUrldate}
\bibliography{Mendeley,local}

\end{document}